\documentclass[twocolumn,english,twocolumns]{revtex4}
\usepackage[T1]{fontenc}
\usepackage[latin9]{inputenc}
\usepackage{amsmath}
\usepackage{amssymb}
\usepackage{graphicx}

\makeatletter
\@ifundefined{textcolor}{}
{%
 \definecolor{BLACK}{gray}{0}
 \definecolor{WHITE}{gray}{1}
 \definecolor{RED}{rgb}{1,0,0}
 \definecolor{GREEN}{rgb}{0,1,0}
 \definecolor{BLUE}{rgb}{0,0,1}
 \definecolor{CYAN}{cmyk}{1,0,0,0}
 \definecolor{MAGENTA}{cmyk}{0,1,0,0}
 \definecolor{YELLOW}{cmyk}{0,0,1,0}
}

\usepackage{babel}

\makeatother

\usepackage{babel}
\begin{document}

\title{Validity of single-channel model for a spin-orbit coupled atomic
Fermi gas near Feshbach resonances}

\author{Jing-Xin Cui$^{1,2}$, Xia-Ji Liu$^{2}$, Gui Lu Long$^{1,3}$, and
Hui Hu$^{2}$}

\email{hhu@swin.edu.au}

\selectlanguage{english}%

\affiliation{$^{1}$Department of Physics, Tsinghua University, Beijing 100084,
China\\
$^{2}$ARC Centre of Excellence for Quantum-Atom Optics, Centre for
Atom Optics and Ultrafast Spectroscopy, Swinburne University of Technology,
Melbourne 3122, Australia\\
$^{3}$Tsinghua National Laboratory for Information Science and Technology,
Tsinghua University, Beijing 100084, China}

\date{\today}
\begin{abstract}
We theoretically investigate a Rashba spin-orbit coupled Fermi gas
near Feshbach resonances, by using mean-field theory and a two-channel
model that takes into account explicitly Feshbach molecules in the
close channel. In the absence of spin-orbit coupling, when the channel
coupling $g$ between the closed and open channels is strong, it is
widely accepted that the two-channel model is equivalent to a single-channel
model that excludes Feshbach molecules. This is the so-called broad
resonance limit, which is well-satisfied by ultracold atomic Fermi
gases of $^{6}$Li atoms and $^{40}$K atoms in current experiments.
Here, with Rashba spin-orbit coupling we find that the condition for
equivalence becomes much more stringent. As a result, the single-channel
model may already be insufficient to describe properly an atomic Fermi
gas of $^{40}$K atoms at a moderate spin-orbit coupling. We determine
a characteristic channel coupling strength $g_{c}$ as a function
of the spin-orbit coupling strength, above which the single-channel
and two-channel models are approximately equivalent. We also find
that for narrow resonance with small channel coupling, the pairing
gap and molecular fraction is strongly suppressed by SO coupling.
Our results can be readily tested in $^{40}$K atoms by using optical
molecular spectroscopy. 
\end{abstract}


\maketitle

\section{Introduction}

As a realization of non-abelian gauge fields in neutral cold atoms
\cite{Lin2011,Williams2012,Zhang2012,Wang2012,Cheuk2012}, spin-orbit
(SO) coupled atomic gases have attracted a lot of attentions in recent
years. The SO coupled bosonic gas of $^{87}\text{{Rb}}$ atoms was
first achieved by Spielman's group at National Institute of Standards
and Technology (NIST) in early 2011 \cite{Lin2011}. The SO coupled
atomic Fermi gas has also been realized most recently at Shanxi University
\cite{Wang2012} and at Massachusetts Institute of Technology (MIT)
\cite{Cheuk2012} with $^{40}\text{K}$ and $^{6}\text{Li atoms}$,
respectively. These novel atomic gases have many interesting properties
inherent to spin-orbit coupling, and have potential applications in
future quantum technology. A well-known example is the emulation of
the long-sought topological superfluids and Majorana fermions \cite{Zhang2008,Liu2012a,Liu2012b},
which lie at the heart of topological quantum information and computation
\cite{Hasan2010,Qi2011}.

Most of previous theoretical studies on SO coupled Fermi gases are
based on a single-channel model \cite{Zhang2008,JPV2011a,JPV2011b,Hu2011,Yu2011,Gong2011,Yi2011,Zhou2011,Sau2011,Iskin2011,Ozawa2011,Zhou2012,Han2012,Seo2012,DellAnna2011,He2012a,He2012b,Jiang2011,Liu2012c,Yang2012,Zhang2012a,Zhang2012b}.
In this model, the interaction between atoms is described by a single
parameter, i.e., the \textit{s}-wave scattering length $a_{s}$. The
scattering length can experimentally be tuned by using Feshbach resonances.
As a result, in the absence of SO coupling, the Fermi gas can cross
from a Bose-Einstein condensate (BEC) over to a Bardeen-Cooper-Schrieffer
(BCS) superfluid \cite{Giorgini2008}, when the scattering length
changes from positive to negative values. With SO coupling, the picture
of BEC-BCS crossover may be qualitatively altered. For example, for
a particular Rashba-type SO coupling, a new two-body bound state -
referred to as rashbon - is formed \cite{JPV2011a,JPV2011b,Hu2011,Yu2011}.
By increasing the SO coupling strength, the system may change from
a BCS superfluid to a BEC of rashbons, even on the BCS side with a
negative scattering $a_{s}<0$ \cite{JPV2011b,Hu2011,Yu2011}. The
pairing gap of this system will be significantly enhanced due to the
increased density of state at the Fermi surface \cite{Yu2011}. An
anisotropic superfluid due to the Rashba SO coupling has also been
predicted \cite{Hu2011}.

A more realistic and complete description of ultracold atomic Fermi
gases near Feshbach resonances, however, should be the two-channel
model, which includes both atoms in the open channel and Feshbach
molecules in the closed channel \cite{Holland2001}. In this model,
in addition to the background \textit{s}-wave scattering length $a_{bg}$
between atoms, two other parameters are used in order to fully describe
the interaction. These are the detuning energy of Feshbach molecules
$\nu$ and the channel coupling strength $g$ between molecules and
Fermi atoms. Therefore, the interaction of the system consists of
two parts. The non-resonant part is the contact interaction between
atoms with the strength determined by the background scattering length,
while the resonant interaction is induced by the coupling between
molecules and atoms. Near Feshbach resonances without SO coupling,
it is known that the single-channel and two-channel models are essentially
equivalent when the channel coupling strength $g$ is large enough
\cite{Diener2004,Liu2005}. This is the so-called broad resonance
condition, satisfied by the Fermi gases of $^{40}$K and $^{6}$Li
atoms, which are so far the two main systems used in the cold-atom
laboratory.

In this paper, we aim to examine the equivalence of the single-channel
and two-channel models for a Rashba SO coupled Fermi gas near Feshbach
resonances. This is by no means obvious, as fermionic pairing is notably
affected by SO coupling at the BEC-BCS crossover. We use mean-field
theory and focus on the most interesting resonant limit. Our results
show that in the presence of SO coupling, the broad resonance condition
is much more difficult to achieve. As a result, for an ultracold atomic
Fermi gas of $^{40}$K atoms, which is known to be well described
by the single-channel model without SO coupling, we may have to use
a two-channel model already at a moderate SO coupling strength.

Our paper is organized as follows. In the next section (Sec. II),
we introduce the model Hamiltonian. In Sec. III, we diagonalize the
Hamiltonian by using mean-field theory to obtain the grand thermodynamic
potential and solve the resulting coupled mean-field equations. In
Sec. IV, we discuss the equivalence between the single-channel and
two-channel models. In Sec. V, we show how to test experimentally
the difference between the two models, by using optical molecular
spectroscopy \cite{Partridge2005}. Finally, we summarize in Sec.
VI.

\section{Model Hamiltonian}

We consider a three-dimensional (3D) resonantly-interacting atomic
Fermi gas with Rashba-type SO coupling, described by the two-channel
model Hamiltonian, 
\begin{equation}
{\cal H}={\cal H}_{SO}+{\cal H}_{m}+{\cal H}_{I}\text{,}\label{totHami}
\end{equation}
 where ${\cal H}_{SO}$ and ${\cal H}_{m}$ stand for the non-interacting
Hamiltonian of SO coupled atoms in the open channel and of Feshbach
molecules in the closed channel, respectively. The interaction Hamiltonian
${\cal H}_{I}={\cal H}_{am}+{\cal H}_{aa}$ includes both the atom-molecule
coupling between the two channels (${\cal H}_{am}$) and the background
interaction between open-channel atoms (${\cal H}_{aa}$).

For atoms, we take the following single-particle Rashba SO Hamiltonian,

\begin{equation}
{\cal H}_{SO}=\frac{\hbar^{2}{\bf k}^{2}}{2m}+\frac{\hbar^{2}}{2m}\lambda{\bf k}_{\perp}\cdot{\bf \sigma}_{\perp},
\end{equation}
 where ${\bf k}_{\perp}\equiv(k_{x},k_{y})$ and ${\bf \sigma}_{\perp}\equiv(\sigma_{x},\sigma_{y})$
are respectively the in-plane momentum and in-plane Pauli matrix,
and $\lambda$ is the Rashba SO coupling strength. Note that, the
standard representation of the Rashba SO coupling is given by $\lambda(k_{y}\sigma_{x}-k_{x}\sigma_{y})$
\cite{Hu2011}. Here, for convenience we have performed a spin-rotation
to rewrite the Rashba term into a slightly different but fully equivalent
form $\lambda(k_{x}\sigma_{x}+k_{y}\sigma_{y})$ \cite{Yu2011}. In
the second quantized form,

\begin{equation}
{\cal H}_{SO}=\sum_{{\bf k}\sigma}\epsilon_{{\bf k}}a_{{\bf k}\sigma}^{\dagger}a_{{\bf k}\sigma}+\frac{\hbar^{2}\lambda k_{\bot}}{2m}\left(e^{-i\varphi_{{\bf k}}}a_{{\bf k}\uparrow}^{\dagger}a_{{\bf k}\downarrow}+\text{H.c.}\right),
\end{equation}
 where $a_{{\bf k}\sigma}^{\dagger}$ is the creation operator for
atoms with momentum ${\bf k}$ in the spin state $\sigma$, $\epsilon_{{\bf k}}\equiv\hbar^{2}{\bf k}^{2}/(2m)$
and $\varphi_{{\bf k}}\equiv\arg(k_{x}+ik_{y})$. To diagonalize this
single-particle Hamiltonian, we introduce the field operators in the
helicity basis labeled by ``$\pm$'',

\begin{equation}
\left(\begin{array}{c}
h_{{\bf k}+}\\
h_{{\bf k}-}
\end{array}\right)=\frac{1}{\sqrt{2}}\left(\begin{array}{cc}
1 & e^{-i\varphi_{{\bf k}}}\\
e^{i\varphi_{{\bf k}}} & -1
\end{array}\right)\left(\begin{array}{c}
a_{{\bf k}\uparrow}\\
a_{{\bf k}\downarrow}
\end{array}\right),
\end{equation}
 with which the single-particle Rashba Hamiltonian becomes diagonal,

\begin{equation}
{\cal H}_{SO}=\sum_{{\bf k}}\left(\epsilon_{{\bf k}+}h_{{\bf k}+}^{\dagger}h_{{\bf k}+}+\epsilon_{{\bf k}-}h_{{\bf k}-}^{\dagger}h_{{\bf k}-}\right).
\end{equation}
Note that, in the helicity basis, the single-particle dispersion relation
now breaks into two branches: $\epsilon_{{\bf k}+}=\epsilon_{{\bf k}}+\hbar^{2}\lambda k_{\bot}/(2m)$
for the upper branch and $\epsilon_{{\bf k}-}=\epsilon_{{\bf k}}-\hbar^{2}\lambda k_{\bot}/(2m)$
for the lower branch. To describe Feshbach molecules, we use annihilation
operators $b_{{\bf q}}$. The energy of molecules is denoted as $2\nu$,
which after renormalization \cite{Holland2001} is related to the
detuning energy from threshold of Feshbach resonance $B_{0}$, i.e.,
$2\nu_{0}=\Delta\mu(B-B_{0})$, where $\Delta\mu\equiv2\mu_{a}-\mu_{m}$
is the magnetic moment difference between the atomic ($2\mu_{a}$)
and bound molecular state ($\mu_{m}$) \cite{Holland2001}. The Hamiltonian
of Feshbach molecules may be written as,

\begin{equation}
{\cal H}_{m}=2\nu\sum_{{\bf q}}b_{{\bf q}}^{\dagger}b_{{\bf q}}.
\end{equation}
 Finally, the interaction Hamiltonian is given by ${\cal H}_{I}={\cal H}_{aa}+{\cal H}_{am}$,
where

\begin{equation}
{\cal H}_{aa}=U_{bg}\sum_{{\bf kk}^{\prime}{\bf q}}a_{{\bf q}/2+{\bf k}\uparrow}^{\dagger}a_{{\bf q}/2-{\bf k}\downarrow}^{\dagger}a_{{\bf q}/2-{\bf k}^{\prime}\downarrow}a_{{\bf q}/2+{\bf k}^{\prime}\uparrow}
\end{equation}
 is the non-resonant interaction between atoms, with strength given
by the background \textit{s}-wave scattering length after renormalization
\cite{Holland2001}, $U_{bg}=4\pi\hbar^{2}a_{bg}/m$, and 
\begin{equation}
{\cal H}_{am}=g\sum_{{\bf kq}}\left[b_{{\bf q}}^{\dagger}a_{{\bf q}/2+{\bf k}\uparrow}a_{{\bf q}/2-{\bf k}\downarrow}+\text{H.c.}\right]
\end{equation}
 is the resonant interaction between atoms and molecules, with strength
parameterized by $g$. After renormalization, the magnitude of the
channel coupling strength $g$ is related to the width of the Feshbach
resonance $W$, i.e., $g\equiv\sqrt{\Delta\mu WU_{bg}}$. We note
that, in the two-channel model one may define an effective \textit{s}-wave
length \cite{Liu2005}, 
\begin{equation}
a_{s}=a_{bg}\left(1-\frac{W}{B-B_{0}}\right)=a_{bg}-\frac{g^{2}}{2\nu_{0}}\frac{m}{4\pi\hbar^{2}}.\label{aseff}
\end{equation}

\section{Mean field theory}

We use the standard mean-field theory to solve the two-channel model
Eq. (\ref{totHami}), by assuming that all the molecules and Cooper
pairs condense into the zero-momentum state. Thus, we set ${\bf q}=0$
in the interaction Hamiltonian ${\cal H}_{I}$. Here, we have excluded
the possibility of an inhomogeneous superfluid phase (i.e., ${\bf q}\neq0$),
which may exist in the presence of an in-plane Zeeman-field \cite{Zheng2012}.
This is consistent with the two-body calculation \cite{Hu2011,Jiang2011,Dong2012}
that the ground state of two particles in our Hamiltonian always has
zero center-of-mass momentum. Following the procedure in Ref. \cite{Holland2001},
we introduce the following field parameters:

\begin{eqnarray}
\phi_{m} & = & \left\langle b_{0}\right\rangle ,\\
p & = & \sum_{{\bf k}}\left\langle a_{{\bf k}\uparrow}a_{-{\bf k}\downarrow}\right\rangle ,\\
f & = & \sum_{{\bf k}}\left\langle a_{{\bf k}\uparrow}^{\dagger}a_{{\bf k}\uparrow}\right\rangle =\sum_{{\bf k}}\left\langle a_{{\bf k}\downarrow}^{\dagger}a_{{\bf k}\downarrow}\right\rangle ,
\end{eqnarray}
 where $\phi_{m}$ is the molecular field in the closed channel, $p$
is the pairing field, and $f$ is half of the number of fermionic
atoms in the open channel. The interaction Hamiltonian ${\cal H}_{aa}$
can therefore be written as 
\begin{equation}
\frac{{\cal H}_{aa}}{U_{bg}}\simeq\sum_{{\bf k}\sigma}fa_{{\bf k}\sigma}^{\dagger}a_{{\bf k}\sigma}-\sum_{{\bf k}}\left(pa_{{\bf k}\uparrow}^{\dagger}a_{-{\bf k}\downarrow}^{\dagger}+\text{H.c.}\right)-|p|^{2}-f^{2}.
\end{equation}
 Similarly, we approximate ${\cal H}_{am}$ as

\begin{equation}
{\cal H}_{am}\simeq-g\sum_{{\bf k}}\left(\phi_{m}a_{{\bf k}\uparrow}^{\dagger}a_{-{\bf k}\downarrow}^{\dagger}+\text{H.c.}\right).
\end{equation}
 Thus, within mean-field the total Hamiltonian is given by,

\begin{eqnarray}
{\cal H} & = & -U_{bg}\left(|p|^{2}+f^{2}\right)+\sum_{{\bf k}\tau}\epsilon_{{\bf k}\tau}h_{{\bf k}\tau}^{\dagger}h_{{\bf k}\tau}+\sum_{{\bf k}\sigma}U_{bg}fa_{{\bf k}\sigma}^{\dagger}a_{{\bf k}\sigma}\nonumber \\
 &  & +2\nu|\phi_{m}|^{2}-\sum_{{\bf k}}\left[\left(U_{bg}p+g\phi_{m}\right)a_{{\bf k}\uparrow}^{\dagger}a_{-{\bf k}\downarrow}^{\dagger}+\text{H.c.}\right],
\end{eqnarray}
 where $\tau\equiv\pm$ is the index of helicity branch. By defining
an order parameter $\Delta=-(U_{bg}p+g\phi_{m})$ and rewriting all
the field operators in the helicity basis, the total mean-field Hamiltonian
becomes 
\begin{eqnarray}
{\cal H} & = & -U_{bg}\left(|p|^{2}+f^{2}\right)+2\nu|\phi_{m}|^{2}+\sum_{{\bf k}\tau}\left(\epsilon_{{\bf k}\tau}+U_{bg}f\right)h_{{\bf k}\tau}^{\dagger}h_{{\bf k}\tau}\nonumber \\
 &  & -\frac{\Delta}{2}\sum_{{\bf k}}\left[e^{-i\varphi_{{\bf k}}}h_{{\bf k}+}^{\dagger}h_{-{\bf k}+}^{\dagger}+e^{i\varphi_{{\bf k}}}h_{{\bf k}-}^{\dagger}h_{-{\bf k}-}^{\dagger}+\text{H.c.}\right].
\end{eqnarray}
 To determine the variational field parameters ($\phi_{m}$, $p$,
and $f$), we diagonalize ${\cal K}={\cal H}-\mu{\cal N}$ by using
Bogoliubov transformation and calculate the grand thermodynamic potential
$\Omega$. Here, ${\cal N}\equiv\sum_{{\bf k}\sigma}a_{{\bf k}\sigma}^{\dagger}a_{{\bf k}\sigma}+2\sum_{{\bf k}}b_{{\bf k}}^{\dagger}b_{{\bf k}}$
is the operator of total number of atoms and $\mu$ is the chemical
potential. Using the field operators for Bogoliubov quasiparticles
\cite{Holland2001}, $\alpha_{{\bf k}+}$ and $\alpha_{{\bf k}-}$,
${\cal K}$ takes the diagonal form, 
\begin{eqnarray}
{\cal K} & = & \sum_{{\bf k}\tau}E_{{\bf k}\tau}\alpha_{{\bf k}\tau}^{\dagger}\alpha_{{\bf k}\tau}-U_{bg}\left(|p|^{2}+f^{2}\right)+2\left(\nu-\mu\right)|\phi_{m}|^{2}\nonumber \\
 &  & +\sum_{{\bf k}}\left[\left(\xi_{\mathbf{k}}+U_{bg}f\right)-\frac{E_{{\bf k}+}+E_{{\bf k}-}}{2}\right],
\end{eqnarray}
 where $\xi_{\mathbf{k}}\equiv\epsilon_{{\bf k}}-\mu$ and the energies
of Bogoliubov quasiparticles $E_{{\bf k}\pm}$ are given by, 
\begin{equation}
E_{{\bf k}\pm}=\sqrt{\left(\xi_{\mathbf{k}}\pm\frac{\hbar^{2}\lambda k_{\bot}}{2m}+U_{bg}f\right)^{2}+|\Delta|^{2}}.
\end{equation}
 At temperature $T$, it is straightforward to write down the grand
thermodynamic potential, 
\begin{eqnarray}
\Omega & = & \sum_{{\bf k}}\left[\xi_{\mathbf{k}}+U_{bg}f-\frac{E_{{\bf k}+}+E_{{\bf k}-}}{2}\right]-U_{bg}\left(|p|^{2}+f^{2}\right)\nonumber \\
 &  & +2\left(\nu-\mu\right)|\phi_{m}|^{2}-k_{B}T\sum_{{\bf k}\tau}\ln\left[1+e^{-\frac{E_{{\bf k}\tau}}{k_{B}T}}\right].
\end{eqnarray}
 The field parameters ($\phi_{m}$, $p$, and $f$) must satisfy the
coupled self-consistent equations, $\partial\Omega/\partial f=0$,
$\partial\Omega/\partial p=0$, and $\partial\Omega/\partial\phi_{m}=0$.
Furthermore, the chemical potential is determined by the total number
of atoms $N$, i.e., 
\begin{equation}
N=-\frac{\partial\Omega}{\partial\mu}=2f+2\phi_{m}^{2}.
\end{equation}
 These four coupled equations can be solved to obtain the pairing
order parameter $\Delta=-(U_{bg}p+g\phi_{m})$ and chemical potential
$\mu$.

\section{Results and discussions\label{sec:thermodynamics}}

To clearly contrast the two-channel model with single-channel model,
we focus on the resonant limit and neglect the back-ground interaction.
By setting $U_{bg}=0$ (and therefore $\Delta=-g\phi_{m}$) and renormalizing
the energy of molecules $\nu$ by using \cite{Holland2001},

\begin{equation}
2\nu=2\nu_{0}+\sum_{{\bf k}}\frac{g^{2}}{2\epsilon_{{\bf k}}},
\end{equation}
 we obtain the coupled gap equation and number equation in the two-channel
model ($\tau\equiv\pm$), 
\begin{eqnarray}
\frac{2(\nu_{0}-\mu)}{g^{2}} & = & \sum_{{\bf k}}\left[\sum_{\tau}\frac{1/2-n_{{\bf k}\tau}}{2E_{{\bf k}\tau}}-\frac{1}{2\epsilon_{{\bf k}}}\right],\label{gapEq2c}\\
N-\frac{2\Delta^{2}}{g^{2}} & = & \sum_{{\bf k}}\left[1-\sum_{\tau}\left(\frac{1}{2}-n_{{\bf k}\tau}\right)\frac{\epsilon_{{\bf k}\tau}-\mu}{E_{{\bf k}\tau}}\right],\label{numEq2c}
\end{eqnarray}
 where $n_{{\bf k}\pm}\equiv1/(e^{E_{{\bf k}\pm}/k_{B}T}+1)$ is the
Fermi-Dirac distribution function. In contrast, the gap and number
equations in the single-channel model are given by \cite{Hu2011},

\begin{eqnarray}
-\frac{m}{4\pi\hbar^{2}a_{s}} & = & \sum_{{\bf k}}\left[\sum_{\tau}\frac{1/2-n_{{\bf k}\tau}}{2E_{{\bf k}\tau}}-\frac{1}{2\epsilon_{{\bf k}}}\right],\\
N & = & \sum_{{\bf k}}\left[1-\sum_{\tau}\left(\frac{1}{2}-n_{{\bf k}\tau}\right)\frac{\epsilon_{{\bf k}\tau}-\mu}{E_{{\bf k}\tau}}\right],
\end{eqnarray}
 respectively. By recalling from Eq. (\ref{aseff}) that the effective
\textit{s}-wave scattering length in the two-channel model is $-4\pi\hbar^{2}a_{s}/m=g^{2}/(2\nu_{0})$,
it is clear that the gap and number equations in both models have
the same structure. However, additional terms, $-2\mu/g^{2}$ and
$-2\Delta^{2}/g^{2}$, appear in the two-channel gap and number equations,
respectively. For a finite SO coupling constant $\lambda$, if $g\rightarrow\infty$,
$2\mu/g^{2}$ and $2\Delta^{2}/g^{2}$ go to zero. Then, the equations
of the two models become exactly the same. This is the same as the
situation without SO coupling. In other words, the single-channel
model and two-channel model coincide with each other in the broad
resonant limit, as they should be. However, for a finite channel coupling
strength $g$, if $\lambda$ is sufficiently large, deep two-body
bound state (i.e., rashbon) appears, with a divergent chemical potential
($\mu\rightarrow-\infty$; see Fig. 1(b) below). Thus, we can not
neglect $2\mu/g^{2}$ in Eq. (\ref{gapEq2c}) anymore and the two
models are no longer equivalent. In this strong SO coupling limit,
we anticipate a qualitative difference between the single-channel
and two-channel models.

Let us turn to detailed numerical calculations. For simplicity, we
consider the case in which the temperature is zero and the system
is exactly at Feshbach resonance ($\nu_{0}=0$ and $a_{s}^{-1}=0$).
To characterize the width of Feshbach resonances, we introduce a {\em
dimensionless} channel coupling constant, 
\begin{equation}
g_{0}=\frac{2m}{\hbar^{2}k_{F}^{1/2}}g,
\end{equation}
 where $k_{F}=(3\pi^{2}N/V)^{1/3}$ is the Fermi wavelength. We take
the energy and length in the units of the Fermi energy $E_{F}=\hbar^{2}k_{F}^{2}/2m$
and the inverse Fermi wavelength $k_{F}^{-1}$, respectively. The
SO coupling strength is measured by the dimensionless parameter $\lambda/k_{F}$.
In experiments, the typical magnitude of SO coupling strength is at
the order of the Fermi wavelength, i.e., $\lambda=O(k_{F})$.

\begin{figure}
\includegraphics[width=1\columnwidth,height=6cm]{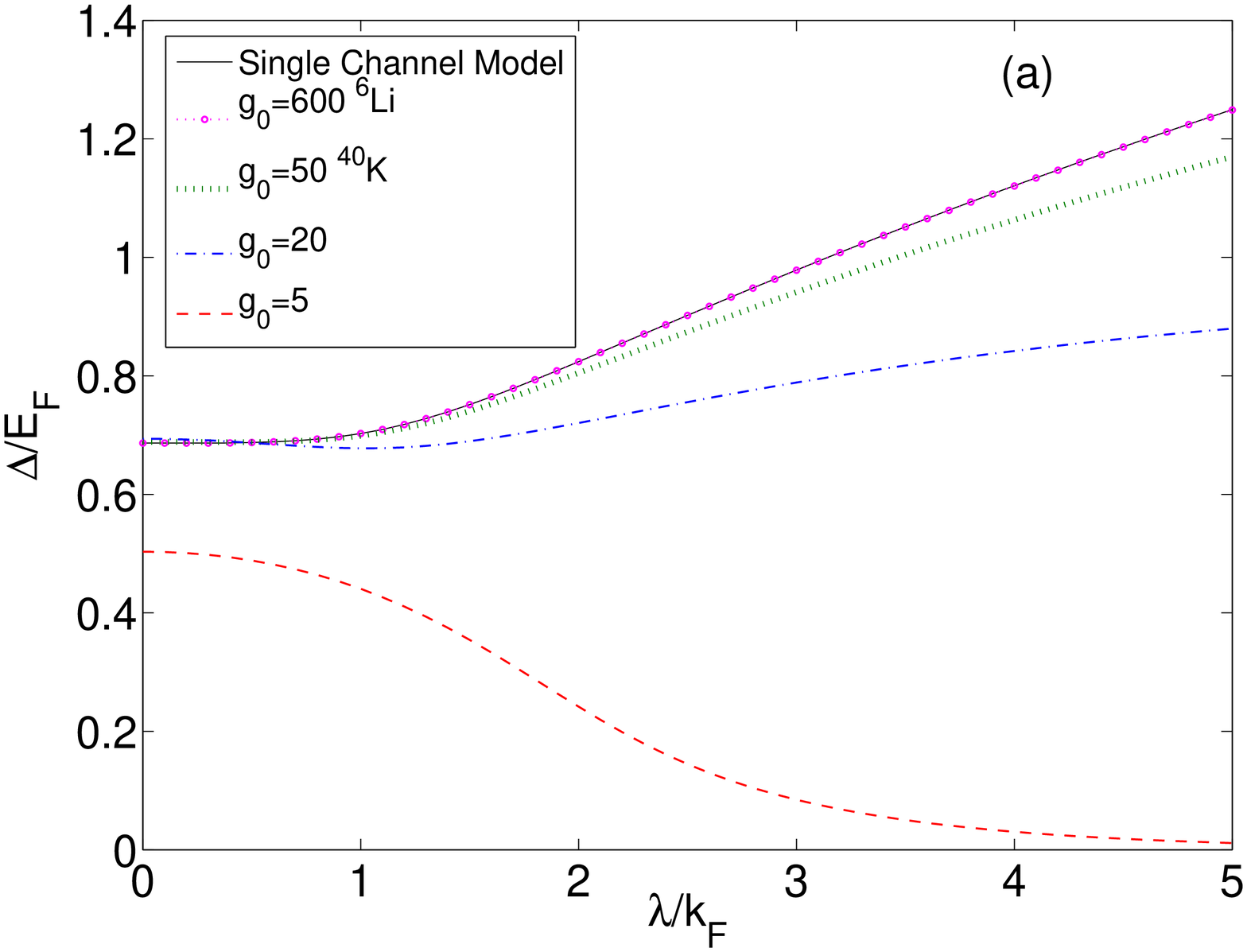} \includegraphics[width=1\columnwidth,height=6cm]{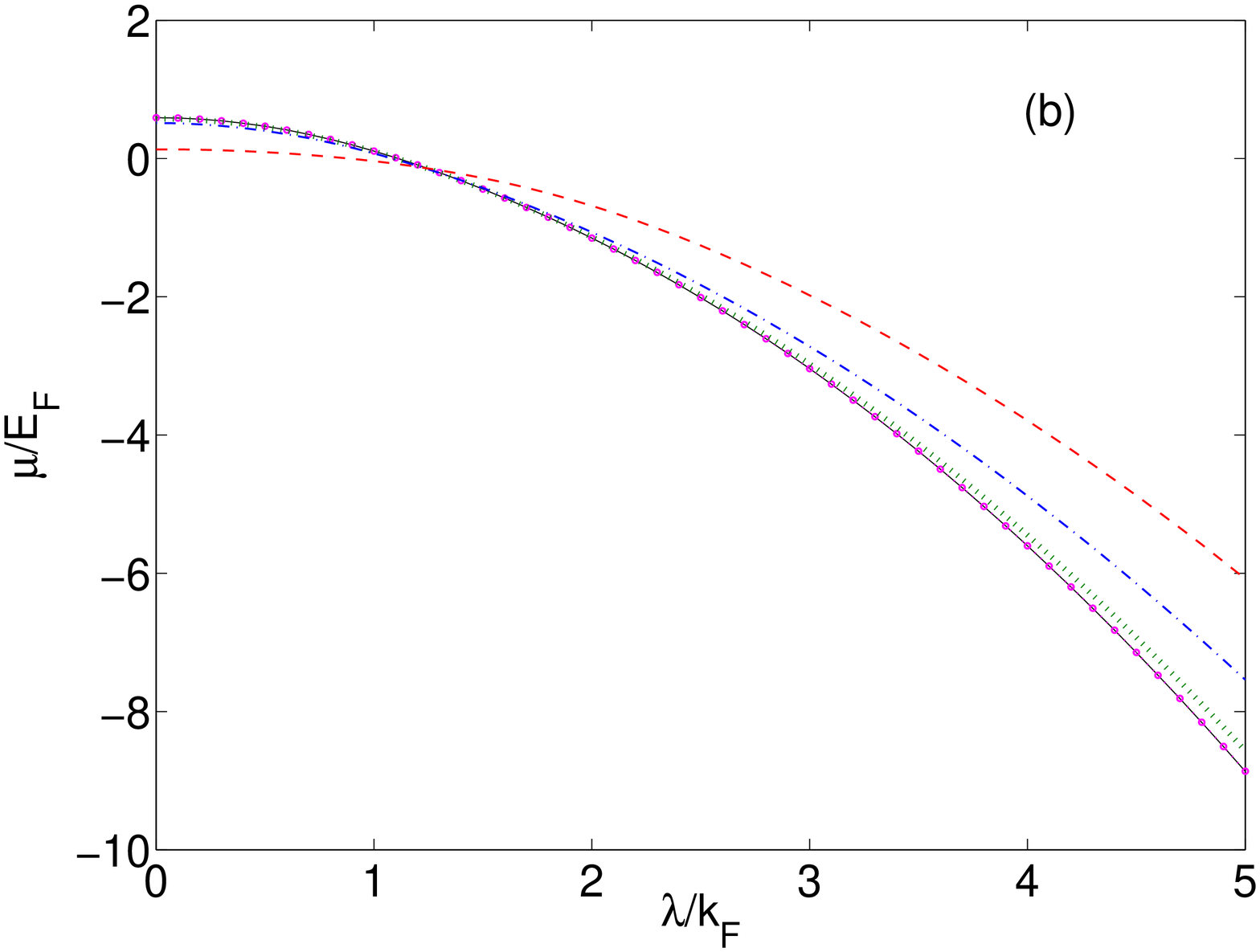}
\caption{The pairing gap $\Delta$ (a) and chemical potential $\mu$ (b) as
a function of the Rashba SO coupling strength $\lambda$ for different
atom-molecule coupling $g_{0}$ or resonance width at zero temperature
and at Feshbach resonance.}
\end{figure}

Fig. 1 reports the evolution of the pairing gap (a) and chemical potential
(b) with decreasing the resonance width. For comparison, the prediction
of single-channel model is also shown by solid lines. As seen from
Fig. 1(a), for systems with small SO coupling strength, the single-channel
model and two-channel model give the same results when $g_{0}$ is
large enough (i.e., $g_{0}>20$). In the case of $^{6}\text{Li}$
atoms with $g_{0}\simeq600$ (the purple dot line in Fig. 1), we can
not see the difference from the single-channel prediction. This means
that the broad resonance condition is always valid for $^{6}\text{Li}$
atoms. However, for a smaller $g_{0}$, e.g., $^{40}\text{K}$ atoms
with $g_{0}\simeq50$, the pairing gap deviates clearly from the single-channel
prediction at the typical experimental SO coupling strength $\lambda/k_{F}=3$,
although the two models give essentially the same pairing gap in the
absence of SO coupling. With increasing the SO coupling strength,
the difference between the two models becomes more significant. For
even smaller $g_{0}$ (i.e., $g_{0}=5$), it is interesting that the
dependence of the pairing gap on SO coupling strength changes qualitatively.
The pairing gap starts to decrease with increasing SO coupling strength
and vanishes at sufficiently large SO coupling.

This dramatic change is somehow not anticipated, as the pairing gap
is always enhanced by SO coupling in the single-channel model. It
is closely related to anisotropic superfluidity caused by the Rashba
SO coupling. As discussed in Ref. \cite{Hu2011}, due to SO coupling
the fermionic superfluid has mixed singlet and triplet components.
The fraction of triplet pairing grows with increasing the SO coupling
strength. Thus, within the single-channel model, the amplitude of
pairing gap reflects both singlet and triplet pairing strengths, and
increases as the SO coupling increases. In the two-channel model,
however, the most important resonant-interaction Hamiltonian ${\cal H}_{am}$
is of \textit{s}-wave character and hence favors the singlet pairing.
As the triplet pairing is favored by Rashba SO coupling, the resonance
width and SO coupling have opposite effects on the pairing gap and
destruct with each other. The destruction becomes very pronounced
with decreasing the resonance width, leading to a completely suppressed
pairing gap at large SO coupling and narrow resonance width. 

The suppression of pairing gap can also be mathematically understood
from the two-channel gap equation, Eq. (\ref{gapEq2c}), where we
may treat $2(\nu_{0}-\mu)$ as the effective energy detuning of Feshbach
molecules. By increasing the SO coupling, the chemical potential will
diverge to $-\infty$$ $, and hence the effective energy detuning
is pushed up to the BCS limit. As a result, the pairing gap becomes
significantly suppressed.

\begin{figure}
\includegraphics[width=1\columnwidth,height=6cm]{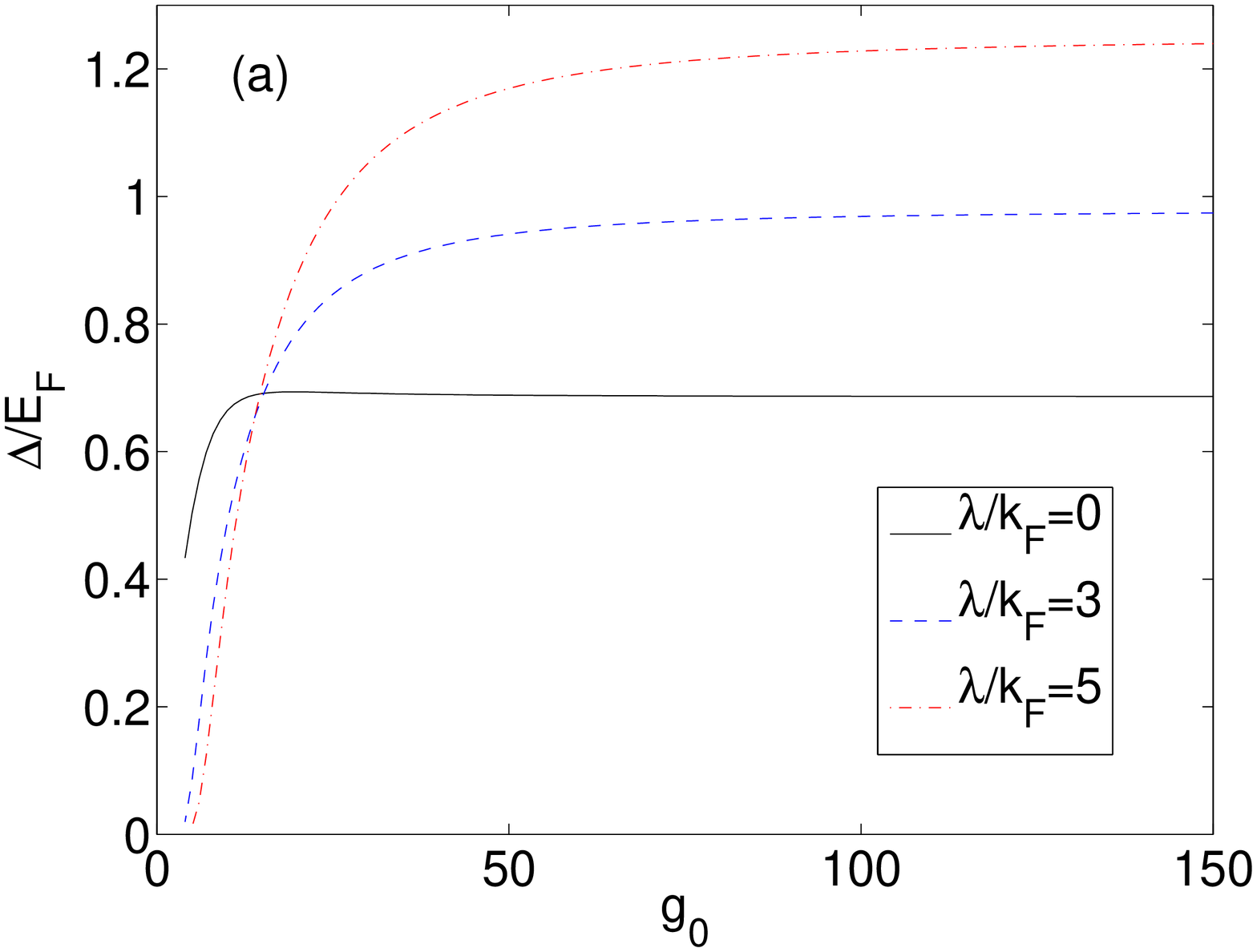} \includegraphics[width=1\columnwidth,height=6cm]{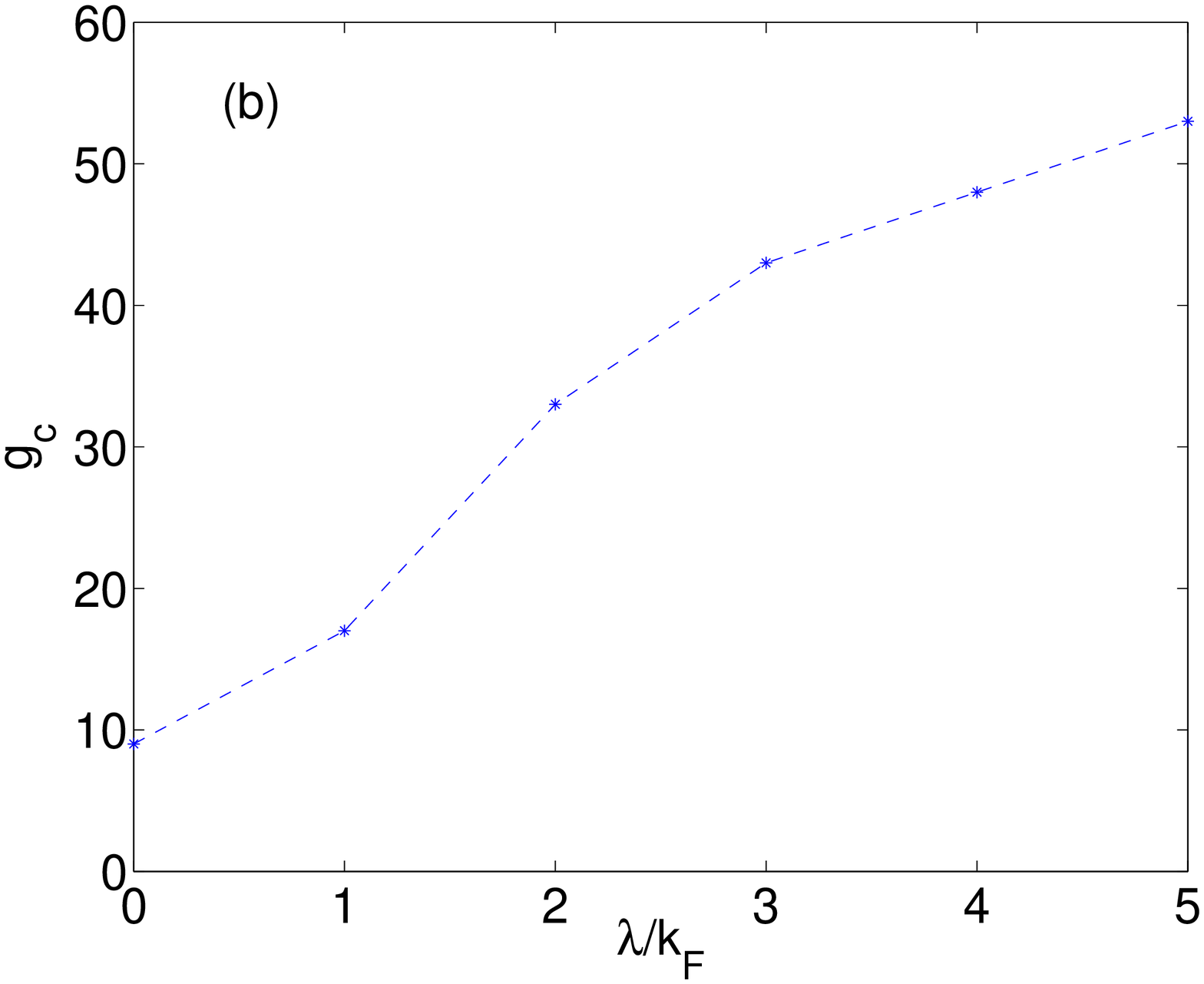}
\caption{(a) The pairing gap $\Delta$ as a function of the resonance width
$g_{0}$ for different Rashba SO coupling strength $\lambda/k_{F}$.
(b) The critical resonance width $g_{0}$ (below which the pairing
gap differs more than 5\% from the prediction of the single-channel
model) for different $\lambda/k_{F}$.}
\end{figure}

Now, it is clear that the broad resonance limit becomes much more
difficult to reach in the presence of Rashba SO coupling. To quantitatively
characterize the broad resonance condition, we show in Fig. 2(a) the
behavior of the pairing gap $\Delta$ as a function of the resonance
width $g_{0}$, for some selected values of $\lambda/k_{F}$. As $g_{0}$
increases, the pairing gap $\Delta$ grows rapidly at first, and then
saturates to the prediction of single-channel model. Quantitatively,
we may define a critical $g_{c}$, in such a way that above $g_{c}$
the relative difference in the pairing gaps predicted by the two models
is less than $5\%$. Fig. 2(b) presents $g_{c}$ as a function of
SO coupling strength. It gives a qualitative phase diagram. Above
$g_{c}$ we may safely use the single-channel model to describe the
Rashba SO coupled atomic Fermi gas near Feshbach resonances. While
below $g_{c}$, the two-channel model must be adopted. For $^{40}\text{K}$
atoms with $g_{0}\simeq50$, we find that the single-channel model
becomes insufficient at a moderate Rashba SO coupling, $\lambda\sim3k_{F}$.

\begin{figure}
\includegraphics[width=1\columnwidth,height=6cm]{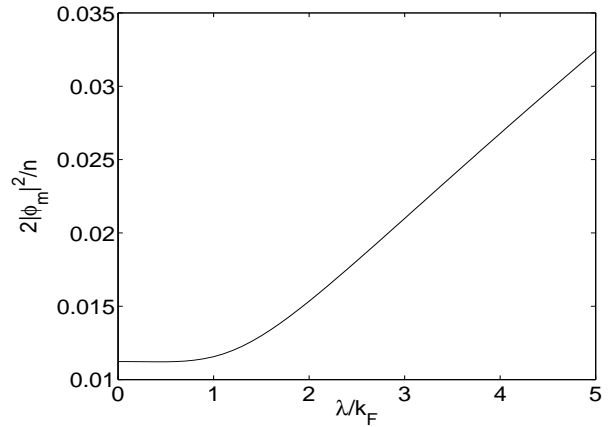} \caption{The molecular fraction $2|\phi_{m}|^{2}/N$ as a function of the Rashba
SO coupling $\lambda/k_{F}$ for $^{40}$K atoms where $g_{0}=50$.}
\end{figure}

\section{Experimental relevance}

To experimentally test our predictions, we consider optical molecular
spectroscopy, which projects Feshbach molecules onto a vibrational
level of an excited molecule. The rate of excitations enables a precise
measurement of the fraction of the closed-channel Feshbach molecules
in the paired state, although the fraction could be extremely small
\cite{Partridge2005}. Near resonance, the paired state may be treated
as dressed molecules \cite{Liu2005,Partridge2005}, 
\begin{equation}
\left|\text{dressed}\right\rangle =e^{i\phi}\sqrt{1-Z_{m}}\left|\text{open}\right\rangle +\sqrt{Z_{m}}\left|\text{closed}\right\rangle ,
\end{equation}
 where $Z_{m}$ can be identified as the component fraction of Feshbach
molecules, i.e., $Z_{m}=2\left|\phi_{m}\right|^{2}/N$. As an concrete
example, in Fig. 3, we show the fraction for $^{40}\text{K}$ atoms
(with $g_{0}\simeq50$) as a function of SO coupling strength. As
the SO coupling increases, the population of Feshbach molecule is
almost flat at first. However, after the coupling reaches a critical
value $\lambda\simeq k_{F}$, it grows very fast.

\begin{figure}
\includegraphics[width=1\columnwidth,height=6cm]{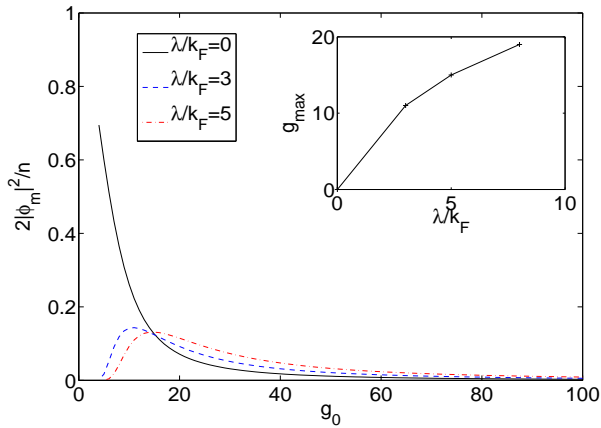} \caption{The molecular fraction as a function of the resonance width $g_{0}$
for different Rashba SO coupling. The inset shows $g_{max}$ (see
text for definition) as a function of the SO coupling strength.}
\end{figure}

The impact of SO coupling on the population of Feshbach molecules
is best seen in Fig. 4, where we present the fraction as a function
of the resonance width at several SO coupling strengths. For a given
non-zero SO coupling, the fraction is a non-monotonic function of
the resonance width. By decreasing $g_{0}$ from the broad resonance
limit, the fraction first grows then drops to zero, as a result of
the competition between SO coupling and resonance width, as mentioned
earlier. In the limit of narrow resonance, the vanishing molecular
fraction is consistent with the suppression in the pairing gap at
large SO coupling, as shown in Fig. 1(a), due to the relation $\Delta=-g\phi_{m}$.
In contrast, in the absence of SO coupling the molecular fraction
increases steadily with decreasing the resonance width. We may define
a characteristic $g_{\max}$ at which the fraction reaches its peak
value. As shown in the inset of Fig. 4, when $\lambda/k_{F}$ is zero,
namely there is no SO coupling, $g_{\max}=0$, and the population
reaches unity in the limit of $g_{0}=0$. As the SO coupling increases,
$g_{\max}$ increases. We emphasize that for small resonance width
$g_{0}$, even a small SO coupling could lead to a strong suppression
of the population of Feshbach molecules.

\section{Summary}

In conclusion, we have investigated a Rashba spin-orbit coupled Fermi
gas near Feshbach resonances, by using a two-channel model. When the
spin-orbit coupling strength is small and Feshbach resonance is broad,
the two-channel model is equivalent to the single-channel model, as
we may anticipate \cite{Diener2004,Liu2005}. However, for a given
resonance width, if the SO coupling strength is sufficiently large,
these two models are no longer equivalent. Moreover, for a narrow
resonance the pairing gap and the fraction of Feshbach molecules are
strongly suppressed by SO coupling. We could test these predictions
by measuring experimentally the molecular fraction using optical molecular
spectroscopy \cite{Partridge2005}. We have characterized quantitatively
the equivalence of the two models by introducing a critical resonance
width, above which the two models are approximately the same. By calculating
the dependence of the critical resonance width on the spin-orbit coupling
strength, we have found that the single channel model may break down
for Rashba spin-orbit coupled $^{40}\text{K}$ atoms at a moderate
spin-orbit coupling strength.

Our results are obtained within mean-field theory, which is known
to provide a qualitative picture of resonantly-interacting atomic
Fermi gases. For quantitative purpose, the crucial pairing fluctuation
must be included. This may be addressed by using many-body \textit{T}-matrix
theories in the future \cite{Liu2005,Liu2006,Hu2006}.
\begin{acknowledgments}
Jing-Xin Cui and Gui Lu Long were supported by the National Natural
Science Foundation of China (Grant No. 11175094), National Basic Research
Program of China (NFRP-China Grant No. 2009CB929402 and No. 2011CB9216002),
and Tsinghua University Initiative Scientific Research Program. Xia-Ji
Liu and Hui Hu were supported by the ARC Discovery Projects (Grant
No. DP0984637 and No. DP0984522) and the NFRP-China (Grant No. 2011CB921502). \end{acknowledgments}

\end{document}